\documentclass[american]{article}
\usepackage[T1]{fontenc}
\usepackage[latin9]{inputenc}
\usepackage[a4paper]{geometry}
\usepackage{babel}
\usepackage{prettyref}
\usepackage{textcomp}
\usepackage{amsmath}
\usepackage{amssymb}
\usepackage{graphicx}
\usepackage[unicode=true,
 bookmarks=true,bookmarksnumbered=false,bookmarksopen=false,
 breaklinks=false,pdfborder={0 0 1},backref=false,colorlinks=false]
 {hyperref}
\hypersetup{pdftitle={Input Synchrony Boosts Correlation Transmission of Pulse-Coupled Units},
 pdfauthor={Matthias Schultze-Kraft, Markus Diesmann, Sonja Grün, Moritz Helias},
 pdfsubject={correlation transmission},
 colorlinks=true,linkcolor=blue,citecolor=blue,urlcolor=blue,filecolor=blue}
\usepackage{breakurl}

\makeatletter

\providecommand{\tabularnewline}{\\}

\newrefformat{fig}{\hyperref[#1]{Fig.~\ref{#1}}}
\newrefformat{tab}{\hyperref[#1]{Table ~\ref{#1}}}
\newrefformat{sec}{\hyperref[#1]{Sec.~\ref{#1}}}
\newrefformat{sub}{\hyperref[#1]{Sec.~\ref{#1}}}
\newrefformat{alg}{\hyperref[#1]{Alg.~\ref{#1}}}
\newrefformat{app}{\hyperref[#1]{App.~\ref{#1}}}

\usepackage[OT1]{fontenc}

\renewenvironment{abstract}
{\noindent{\normalfont\large\textbf{Abstract}}%
\par\vspace{0.5\baselineskip}\noindent}
{\par}

\renewcommand{\@seccntformat}[1]{%
\csname the#1\endcsname\hspace{0.5em}}

\renewcommand{\section}{\@startsection
{section}%
{1}%
{0mm}%
{-\baselineskip}%
{0.5\baselineskip}%
{\normalfont\large\bfseries}}

\renewcommand{\subsection}[1]{\ssubsection{#1.}}
\newcommand{\ssubsection}{\@startsection
{subsection}%
{2}%
{1em}%
{-\baselineskip}%
{-\fontdimen2\font plus -\fontdimen3\font minus -\fontdimen4\font}%
{\normalfont\bfseries}}

\renewcommand{\subsubsection}[1]{\sssubsection{#1.}}
\newcommand{\sssubsection}{\@startsection
{subsubsection}%
{3}%
{1em}%
{-\baselineskip}%
{-\fontdimen2\font plus -\fontdimen3\font minus -\fontdimen4\font}%
{\normalfont\itshape}}

\usepackage{graphics}

\usepackage{footmisc}

\renewcommand{\@makecaption}[2]{%
{\parbox[t]{\linewidth}{%
\normalsize\renewcommand{\baselinestretch}{1.0}\normalsize
\vspace{2mm}
\textbf{#1:} #2
}}}

\usepackage{xspace}

\usepackage{xcolor}
\definecolor{lightgray}{gray}{0.9}

\usepackage{listings}
\makeatother

\begin{document}
\global\long\def\Hz{\;\mathrm{Hz}}
\global\long\def\mV{\;\mathrm{mV}}
\global\long\def\ms{\;\mathrm{ms}}
\global\long\def\exc{\mathrm{exc}}
\global\long\def\inh{\mathrm{inh}}
\global\long\def\inp{\mathrm{in}}
\global\long\def\out{\mathrm{out}}
\global\long\def\Pint{H}
\global\long\def\taum{\tau_{m}}
\global\long\def\Var{\mathrm{Var}}
\global\long\def\Cov{\mathrm{Cov}}
\global\long\def\bc{\bar{c}}
\global\long\def\bnu{\bar{\nu}}
\global\long\def\rhoin{\rho_{\inp}}
\global\long\def\rhoout{\rho_{\out}}
\global\long\def\nuout{\nu_{\out}}
\global\long\def\nuin{\nu_{\mathrm{in}}}
\begin{titlepage}\thispagestyle{empty}\setcounter{page}{0}\pdfbookmark[1]{Title}{TitlePage}

\begin{center}
\textbf{\Large Noise Suppression and Surplus Synchrony by Coincidence
Detection}
\par\end{center}{\Large \par}

\begin{center}
\textbf{Matthias Schultze-Kraft$^{1,2,3}$, Markus Diesmann$^{4,5,7}$,
Sonja Grün$^{4,6,7}$, Moritz Helias$^{4}$}
\par\end{center}

\vspace{3cm}

$^{1}$\parbox[t]{12cm}{Neurotechnology Group\\
Berlin Institute of Technology\\
Berlin\\
Germany}\\[3mm]

$^{2}$\parbox[t]{12cm}{Bernstein Focus: Neurotechnology\\
Berlin\\
Germany}\\[3mm]

$^{3}$\parbox[t]{12cm}{Bernstein Center for Computational Neuroscience\\
Berlin\\
Germany}\\[3mm]

$^{4}$\parbox[t]{12cm}{Institute of Neuroscience and Medicine (INM-6)\\
Computational and Systems Neuroscience\\
Research Center Jülich\\
Jülich\\
Germany}\\[3mm]

$^{5}$\parbox[t]{12cm}{Medical Faculty\\
RWTH Aachen University\\
Aachen\\
Germany}\\[3mm]

$^{6}$\parbox[t]{12cm}{Theoretical Systems Neurobiology\\
RWTH Aachen University\\
Aachen\\
Germany}\\[3mm]

$^{7}$\parbox[t]{12cm}{RIKEN Brain Science Institute\\
Wako City\\
Japan}\\[3mm]\noindent

\noindent\raisebox{11.8cm}[0cm][0cm]{\hspace*{12.5cm}\textbf{}}\vfill{}
\noindent\\
Correspondence to:\hspace{1em}\parbox[t]{11cm}{Matthias Schultze-Kraft\\

\newlength{\myw}\settowidth{\myw}{fax:\ }\makebox[\myw][l]{tel: +49-(0)-30-314-28678}

fax: +49-(0)-30-314-78622\\\
\href{mailto:schultze-kraft@tu-berlin.de}{schultze-kraft@tu-berlin.de}

}\end{titlepage}
\begin{abstract}
The functional significance of correlations between action potentials
of neurons is still a matter of vivid debates. In particular it is
presently unclear how much synchrony is caused by afferent synchronized
events and how much is intrinsic due to the connectivity structure
of cortex. The available analytical approaches based on the diffusion
approximation do not allow to model spike synchrony, preventing a
thorough analysis. Here we theoretically investigate to what extent
common synaptic afferents and synchronized inputs each contribute
to closely time-locked spiking activity of pairs of neurons. We employ
direct simulation and extend earlier analytical methods based on the
diffusion approximation to pulse-coupling, allowing us to introduce
precisely timed correlations in the spiking activity of the synaptic
afferents. We investigate the transmission of correlated synaptic
input currents by pairs of integrate-and-fire model neurons, so that
the same input covariance can be realized by common inputs or by spiking
synchrony. We identify two distinct regimes: In the limit of low correlation
linear perturbation theory accurately determines the correlation transmission
coefficient, which is typically smaller than unity, but increases
sensitively even for weakly synchronous inputs. In the limit of high
afferent correlation, in the presence of synchrony a qualitatively
new picture arises. As the non-linear neuronal response becomes dominant,
the output correlation becomes higher than the total correlation in
the input. This transmission coefficient larger unity is a direct
consequence of non-linear neural processing in the presence of noise,
elucidating how synchrony-coded signals benefit from these generic
properties present in cortical networks.
\end{abstract}

\section*{Author summary}

Whether spike timing conveys information in cortical networks or whether
the firing rate alone is sufficient is a matter of controversial debates,
touching the fundamental question how the brain processes, stores,
and conveys information. If the firing rate alone is the decisive
signal used in the brain, correlations between action potentials are
just an epiphenomenon of high convergence and divergence the cortex'
connectivity, where pairs of neurons share a considerable fraction
of common afferents. Due to the membrane leakage, the small synaptic
amplitudes and the non-linear threshold, nerve cells exhibit lossy
transmission of correlation, if the correlation originates from shared
synaptic inputs. However, the membrane potential of cortical neurons
often displays non-Gaussian fluctuations, a hallmark of synchronized
synaptic afferents. Moreover, synchronously active neurons have been
found to reflect behavior in primates. In this work we therefore contrast
the transmission of correlation due to shared afferents and due to
synchronously arriving synaptic impulses for leaky neuron models.
We not only find that neurons are highly sensitive to synchronous
afferents, but that they are generically able to perform noise suppression
on synchrony coded signals, a computational advantage over rate signals.

\section{Introduction}

Simultaneously recording the activity of multiple neurons provides
a unique tool to observe the activity in the brain. The immediately
arising question of the meaning of the observed correlated activity
between different cells \cite{Perkel67b,Gerstein69} is tightly linked
to the problem how information is represented and processed by the
brain. This problem is matter of an ongoing debate \cite{Cohen11_811}
and has lead to two opposing views. In one view, the high variability
of the neuronal response \cite{Arieli96_1868} to presented stimuli
and the sensitivity of network activity to the exact timing of spikes
\cite{London10_123} suggests that the slowly varying rate of action
potentials carries the information in the cortex. A downstream neuron
can read out the information by pooling a sufficient number of merely
independent stochastic source signals. Correlations between neurons
may either decrease the signal-to-noise ratio of population signals
\cite{Zohary94_140} or they may enhance the information in such population,
depending on the readout mechanism \cite{Shamir01_277}. Correlations
are an unavoidable consequence of cortical connectivity where pairs
of neurons share a considerable amount of common synaptic afferents
\cite{Shadlen98}. Recent works have reported very low average correlations
in cortical networks on long time scales \cite{Ecker10}, explainable
by an active mechanism of decorrelation \cite{Hertz10_427,Renart10_587,Tetzlaff12_4393}.
On top of these correlations inherent to cortex due to its connectivity,
a common and slowly varying stimulus can evoke correlations on a long
time scale.

In the other view, on the contrary, theoretical considerations \cite{Hebb49,Malsburg81,Bienenstock95,Singer95}
argue for the benefit of precisely timed action potentials to convey
and process information by binding elementary representations into
larger percepts. Indeed, in frontal cortex of macaque, correlated
firing has been observed to be modulated in response to behavioral
events, independent of the neurons' firing rate \cite{Riehle97_1950}.
On a fine temporal scale, synchrony of action potentials \cite{Abeles82b,Gruen99_67,Gruen02b}
has been found to dynamically change in time in relation to behavior
in primary visual cortex \cite{Maldonado08_1523} and in motor cortex
\cite{Riehle97_1950,Kilavik09_12653}. The observation that nearby
neurons exclusively show positive correlations suggests common synaptic
afferents to be involved in the modulation of correlations \cite{Vaadia95a}.
In this view, the measure of interest are correlations on a short
temporal scale, often referred to as synchrony.

The role of correlations entails the question whether cortical neurons
operate as integrators or as coincidence detectors \cite{Abeles82b,Koenig96}.
Recent studies have shown that single neurons may operate in both
regimes \cite{Hong12_1413}. If the firing rate is the decisive signal,
integrator properties become important, as neural firing is driven
by the mean input. As activity is modulated by the slowly varying
signal, correlations extend to long time scales due to co-modulation
of the rate. Integrators are thus tailored to the processing of rate
coded signals and they transmit temporal patterns only unreliably.
Coincidence detectors preferentially fire due to synchronously arriving
input. The subthreshold membrane potential fluctuations reflect the
statistics of the summed synaptic input \cite{Lampl99_361}, which
can be used to identify temporally precise repetition of network activity
\cite{Ikegaya04}. A direct probe for the existence of synchronous
activity are the resulting strong deflections due to synchronous arrival
of synaptic impulses. Such non-Gaussian fluctuations have indeed been
observed in auditory cortex in vivo \cite{DeWeese06_12206} and in
the barrel cortex of behaving mice \cite{Poulet08_881}. In this regime,
coincidence detector properties become crucial. Coincidence detectors
are additionally sensitive to stimulus variance \cite{Hong12_1413,Silberberg04_704}
and exhibit correlations arising from precisely timed firing. This
type of correlation is unaffected by firing rate, can encode stimulus
properties independently and moreover arises on short time scales
\cite{Hong12_1413}.

The pivotal role of correlations distinguishing the two opposing views
suggests to ask the following question: Can the experimentally observed
synchrony between the activity of two neurons be explained by common
input or is synchrony in the input required? If common input is sufficient,
synchrony is just a side effect of rate coding. However, if synchrony
in the input is required, this synchrony is likely to propagate information
though the network, as it appears at task-specific times \cite{Riehle97_1950,Maldonado08_1523,Kilavik09_12653}.
A functional interpretation is assigned to synchrony by the picture
of the cell assembly \cite{Hebb49,Malsburg81,Palm90,Singer99_49},
where jointly firing neurons dynamically form a functionally relevant
subnetwork. Due to the local connectivity with high divergence and
convergence, any pair of neurons shares a certain amount of input.
This common input may furthermore exhibit spike synchrony, representing
the coherent activity of the other members of the cell assembly. In
the assembly picture, the synchronous input from peer neurons of the
same assembly is thus considered conveying the signal, while the input
from neurons outside of the assembly is considered as noise \cite{Denker10}.

One particular measure for assessing the transmission of correlation
by a pair of the neurons is the transmission coefficient, i.e. the
ratio of output to input correlation. When studying spiking neuron
models, the synaptic input is typically modeled as Gaussian white
noise, e.g. by applying the diffusion approximation to the leaky integrate-and-fire
model \cite{Brunel00_183}. In the diffusion limit, the transmission
coefficient of a pair of model neurons receiving correlated input
mainly depends on the firing rate of the neurons alone \cite{DeLaRocha07_802,Shea-Brown08}.
For low correlations, linear perturbation theory well describes the
transmission coefficient, which is always below unity, i.e. the output
correlation is bounded by the input correlation, pairs of neurons
always lose correlation. Analytically tractable approximations of
leaky integrate-and-fire neural dynamics have related the low correlation
transmission to the limited memory of the membrane voltage \cite{Rosenbaum10_00116}.
The transmission is lowest if neurons are driven by excitation and
inhibition, when fluctuations dominate the firing. In the mean driven
regime the transmission coefficient can reach unity for integral measures
of correlation \cite{Rosenbaum10_00116}.

Understanding the influence of synchrony among the inputs on the correlation
transmission requires to extend the above mentioned methods, as Gaussian
fluctuating input does not allow to represent individual synaptic
events, not to mention synchrony. Therefore, in this work we introduce
an input model that extends the commonly investigated Gaussian white
noise model. We employ the multiple interaction process (MIP) \cite{Kuhn03_67}
to generate an input ensemble of Poisson spike trains with a predefined
pairwise correlation coefficient. We use these processes containing
spike synchrony as the input common to both neurons and model the
remaining afferents as independent Poisson spike trains. Furthermore,
contrary to studies that measure the integrated output correlation
(count correlation) \cite{DeLaRocha07_802,Shea-Brown08}, we primarily
consider the output correlation on the time scale of milliseconds,
i.e. the type of correlation determined by the coincidence detection
properties of neurons.

In \prettyref{sec:Results} we first introduce the neuron and input
models. In \prettyref{sub:understanding-and-isolating} we study the
impact of input synchrony on the firing properties of a pair of leaky
integrate-and-fire neurons with current based synapses. Isolating
and controlling this impact allows us to separately study the effect
of input synchrony on the one hand and common input on the other hand
on the correlation transmission. In \prettyref{sub:low_corr_limit}
and \prettyref{sub:high_corr_limit} we present a quantitative explanation
of the mechanisms involved in correlation transmission, in the limit
of low and high correlation, respectively, and show how the transmission
coefficient can exceed unity in the latter case. In \prettyref{sec:Discussion}
we summarize our findings in the light of previous research, provide
a simplified model that enables an intuitive understanding and illustrates
the generality of our findings. Finally, we discuss the limitations
of our theory and consider possible further directions.

\section{Results\label{sec:Results}}

The neuronal dynamics considered in this work follows the leaky integrate-and-fire
model, whose membrane potential $V(t)$ obeys the differential equation

\begin{equation}
\taum\frac{dV(t)}{dt}=-(V(t)-V_{0})+\taum s_{\exc}(t)+\taum s_{\inh}(t),\label{eq:diffeq_V_t}
\end{equation}
\[
V(t)\leftarrow V_{r}\;\text{if}\; V(t)>V_{\theta},
\]
where $\tau_{m}$ is the membrane time constant, $V_{0}$ the resting
potential, $V_{\theta}$ the firing threshold, and $V_{r}$ the reset
potential of the neuron. The neuron is driven by excitatory and inhibitory
afferent spike trains $s_{\exc}(t)=w\sum_{j}\delta(t-t_{\exc}^{j})$
and $s_{\inh}(t)=-g\cdot w\sum_{k}\delta(t-t_{\inh}^{k})$ where $w$
is the excitatory synaptic weight and $t_{\exc}^{j}$ and $t_{\inh}^{k}$
are the arrival time points of excitatory and inhibitory synaptic
events, respectively. $s_{\exc/\inh}$ denote the weighted sum of
all afferent excitatory and inhibitory events, respectively. Inhibitory
events are further weighted by the factor $-g$. Each single incoming
excitatory or inhibitory event causes a jump of the membrane potential
by the synaptic weight $w$ or -$gw$, respectively, according to
\prettyref{eq:diffeq_V_t}. Whenever the membrane potential reaches
the threshold $V_{\theta}$ the neuron fires a spike and the membrane
potential is reset to $V_{r}$ after which it is clamped to that voltage
for a refractory period of duration $\tau_{r}$. In the current work
we measure the correlation between two spike trains $s_{i}$ and $s_{j}$
on the time scale $\tau$ as

\begin{equation}
\rho_{\mathrm{out}}^{\tau}=\left\langle \frac{\Cov[n_{i}^{\tau},n_{j}^{\tau}]}{\sqrt{\Var[n_{i}^{\tau}]\Var[n_{j}^{\tau}]}}\right\rangle _{T},\label{eq:spike_count_correlation}
\end{equation}
where $n_{i}^{\tau}$ is the spike count of spike train $s_{i}$ in
a time window $\tau$ and the average $\langle\rangle_{T}$ is performed
over the $T/\tau$ time bins of a stationary trial. Except for \prettyref{sub:low_corr_limit}
where we choose $\tau=100\ms$, throughout our work we choose $\tau=1\ms$
and refer to $\rho_{\out}^{1\ms}$ as output spike synchrony.

\begin{figure}
\begin{centering}
\includegraphics{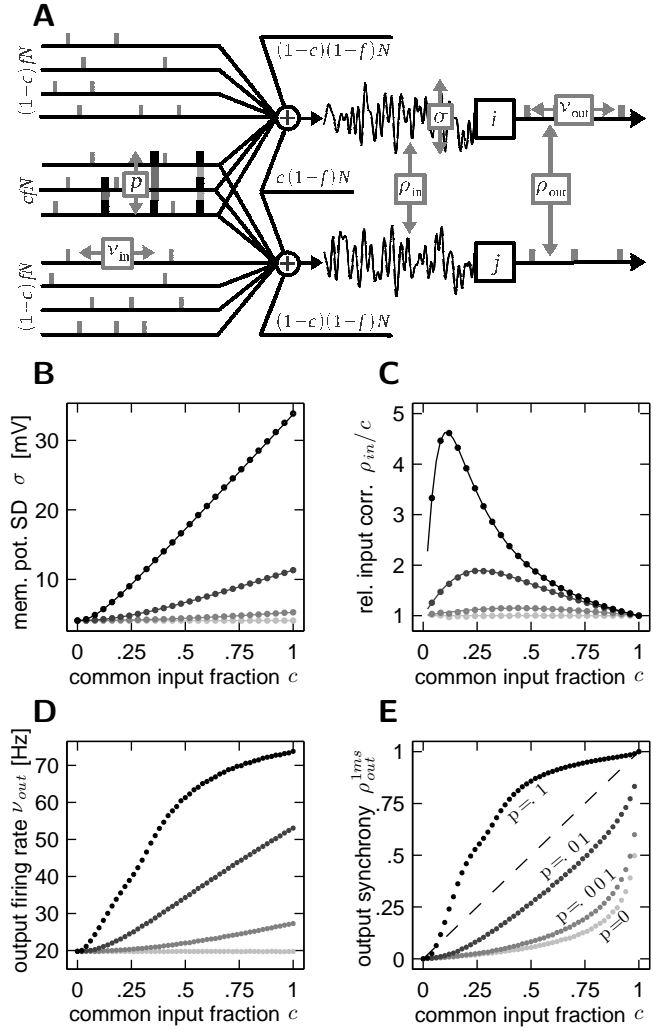} 
\par\end{centering}

\caption{A pair of integrate-and-fire model neurons driven by partially shared
and correlated presynaptic events. \textbf{A} Each of the neurons
$i$ and $j$ receives input from $N$ sources, of which $fN$ are
excitatory and $(1-f)N$ are inhibitory. Both neurons share a fraction
$c$ of their excitatory and inhibitory sources, whereas the fraction
$(1-c)$ is independent for each neuron. Schematically represented
spike trains on the left of the diagram show the excitatory part of
the input, the inhibitory input is only indicated. A single source
emits spike events with a firing rate $\nu_{\inp}$, with marginal
Poisson statistics. Correlated spiking is introduced in the $cfN$
common excitatory sources to both neurons. This pairwise correlation
is realized by means of a multiple interaction process (MIP) \cite{Kuhn03_67}
that yields a correlation coefficient of $p$ between any pairs of
sources. In absence of a threshold, the summed input drives the membrane
potential to a particular working point described by its mean $\mu$
and standard deviation $\sigma$ and the correlation coefficient $\rho_{\inp}=\Cov[V_{i},V_{j}]/(\sigma_{i}\sigma_{j})$
between the free membrane potentials $V_{i}$, $V_{j}$ of both neurons.
In presence of a threshold mean and variance of the membrane potential
determine the output firing rate $\nu_{\out}$ and their correlation
in addition determines the output correlation $\rho_{\out}$, calculated
by \prettyref{eq:spike_count_correlation}. \textbf{B}-\textbf{E}
Direct simulation was performed using different values of common input
fraction $c$ and four fixed values of input spike synchrony $p$
(as denoted in E). Each combination of $c$ and $p$ was simulated
for $100$ seconds, gray coded data points show the average over $50$
independent realizations. Remaining parameters are given in \prettyref{tab:sim_parameters-1}.
Solid lines in B and C are calculated as \prettyref{eq:sigma2} and
\prettyref{eq:rho_in}, respectively. In C, for convenience, $\rhoin$
is normalized by the common input fraction $c$, so that $\rhoin/c=1$
in absence of synchrony ($p=0$). In E, we measure output spike synchrony
$\rho_{\out}^{1\ms}$.\label{fig:1}}
\end{figure}

We investigate the correlation transmission of a pair of neurons considering
the following input scenario. Each neuron receives input from $N$
presynaptic neurons of which $fN$ are excitatory and $(1-f)N$ are
inhibitory. Both neurons share a fraction $c\in\left[0,1\right]$
of their excitatory and inhibitory afferents. \prettyref{fig:1}A
shows a schematic representation of the input to neurons $i=1,2$.
Each source individually obeys Poisson statistics with rate $\nuin$.
We assume that the two neurons under consideration are part of a subnetwork
performing some function. They receive a number $cfN$ common excitatory
inputs from neurons belonging to the same functional unit. We want
to investigate how synchrony can be used to convey information within
the functional subnetwork. We therefore allow synchronization of spikes
to occur in the $cfN$ common excitatory inputs, whereas the $(1-c)fN$
excitatory disjoint afferents are assumed to originate from independent
source networks. The assumption of independence also holds for all
$(1-f)N$ inhibitory sources. Here in particular, we use a multiple
interaction process (MIP) \cite{Kuhn03_67} to model synchronous spike
events among the common, excitatory afferents. In this model each
event of a mother Poisson process of rate $\lambda_{m}$ is copied
independently to any of the $cfN$ child spike trains with probability
$p$, resulting in a pairwise correlation coefficient of $p$ between
two child spike trains. By choosing the rate of the mother spike train
as $\lambda_{m}=\frac{\nuin}{p}$ the rate of a single child spike
train is $\nuin$ and independent of $p$.

\prettyref{fig:1}B, C, D and E show that the amount of pairwise correlations
in the common input has a strong impact on the variance and correlation
of the free membrane potentials ($\sigma^{2},\rhoin$) and therefore
on the output firing rate and output spike synchrony ($\nuout,\rho_{\mathrm{out}}^{1\ms}$).
Let us first consider the case of $p=0$, i.e. the absence of synchronous
events in the input. As expected, the free membrane potential variance
$\sigma^{2}$ remains constant throughout the whole range of $c$,
as does the firing rate $\nu_{\out}$ (\prettyref{fig:1}B and D).
Figure \prettyref{fig:1}C shows the correlation of the free membrane
potential of a neuron pair, normalized by the common input fraction
$c$. As expected, for $p=0$ the input correlation is only determined
by the common input fraction and thus $\rhoin=c$. Hence, the output
synchrony observed for $p=0$ in \prettyref{fig:1}E is solely due
to the correlation caused by common input and describes the often
reported correlation transmission function of the integrate-and-fire
model \cite{DeLaRocha07_802,Shea-Brown08}, where for $0<c<1$ the
output spike synchrony is always well below the identity line, which
is in full agreement with the work of \cite{DeLaRocha07_802}.

Let us now consider the case of $p>0$. In \prettyref{fig:1}B and
D we observe that even small amounts of input synchrony result in
an increased variance of the free membrane potential, which is accompanied
by an increase of the output firing rate. While for weak input synchrony
the increase of $\sigma$ and $\nu_{\out}$ is only moderate, in the
extreme case of strong input synchrony ($p=0.1$) $\sigma$ becomes
almost ten-fold higher and $\nu_{\out}$ increases more than three-fold
compared to the case of $p=0$. \prettyref{fig:1}C shows that input
synchrony also has a strong impact on the correlation between the
free membrane potentials of a neuron pair. For any $p>0$ the input
correlation is most pronounced for high $p$ and in the lower regime
of $c$. Simulation results shown in \prettyref{fig:1}E suggest that
this increase of input correlation is accompanied by an increased
synchrony between the output spikes for $p=0.001$ and $p=0.01$.
For strong input synchrony of $p=0.1$ the output synchrony is always
higher than the input correlation caused solely by the common input,
except near $c=0$ and at $c=1$.

The output firing rates and output spike synchrony shown in figures
\prettyref{fig:1}D and E bear a remarkable resemblance, most notably
for lower values of $c$. Particularly salient is the course of these
quantities for $p=0.1$, which is almost identical over the whole
range of $c$. These observations clearly corroborate findings from
previous studies that have shown an increase of the correlation transmission
of a pair of neurons with the firing rate of the neurons \cite{DeLaRocha07_802,Shea-Brown08}.
Thus, we must presume that a substantial amount of the output synchrony
observed in \prettyref{fig:1}E can be accounted for by the firing
rate increase observed in \prettyref{fig:1}D. Furthermore, as \prettyref{fig:1}C
suggests, for any $p>0$ common input and the synchronous events both
contribute to the correlation between the membrane potentials of a
neuron pair.

\begin{table}
\begin{tabular}{|c|c|c|c|c|c|c|c|c|c|}
\hline 
$N$ & $f$ & $g$ & $w$ & $\nuin$ & $\mu_{0}=I_{0}R_{m}$ & $\tau_{m}$ & $V_{r}$ & $V_{\theta}$ & $\tau_{r}$\tabularnewline
\hline 
\hline 
$4230$ & $0.8$ & $4$ & $0.14\mV$ & $10\; Hz$ & $10\mV$ & $10\ms$ & $0\mV$ & $15\mV$ & $2\ms$\tabularnewline
\hline 
\end{tabular}

\caption{Parameters of the input and LIF neuron used in the simulations}

\label{tab:sim_parameters-1}
\end{table}

\subsection{Understanding and isolating the effect of synchrony\label{sub:understanding-and-isolating}}

These two observations -- the increase of input correlation and output
firing rate induced by input synchrony -- foil our objective to understand
the sole impact of synchronous input events on the correlation transmission
of neurons. In the following we will therefore first provide a quantitative
description of the effect of finite sized presynaptic events on the
membrane potential dynamics and subsequently describe a way to isolate
and control this effect.

The synchronous arrival of $k$ events has a $k$-fold effect on the
voltage due to the linear superposition of synaptic currents. The
total synaptic input can hence be described by a sequence of time
points $t^{j}$ and independent and identically distributed (i.i.d)
random number $w^{j}$ that assume a discrete set of synaptic amplitudes
each with probability $P(w^{j})$. The train of afferent impulses
follows Poisson statistics with some rate $\lambda$. Assuming small
weights $w$ and high, stationary input rate $\lambda$, a Kramers-Moyal
expansion \cite{Brunel00,Risken96,Ricciardi99} can be applied to
\prettyref{eq:diffeq_V_t} to obtain a Fokker-Planck equation for
the membrane potential distribution $p(V,t)$

\begin{align}
\frac{\partial p(V,t)}{\partial t} & =-\frac{\partial}{\partial V}S(V,t)\label{eq:P(V,t)}\\
S(V,t) & =-\frac{\sigma^{2}}{\tau_{m}}\frac{\partial p}{\partial V}(V,t)-\frac{V-\mu}{\tau_{m}}p(V,t).\nonumber 
\end{align}
Only the first two moments $\left\langle w^{j}\right\rangle =\sum_{w^{j}}w^{j}\, P(w^{j})$
and $\langle(w^{j})^{2}\rangle=\sum_{w^{j}}(w^{j})^{2}\, P(w^{j})$
of the amplitude distribution enter the first ($\mu$) and second
($\sigma^{2}$) infinitesimal moments as \cite[cf. Appendix Input-Output Correlation of an Integrate-and-Fire Neuron for a detailed derivation]{Helias08_7} 

\begin{eqnarray}
\mu & = & \lambda\taum\langle w^{j}\rangle+V_{0}\label{eq:infinitesimal_moments}\\
\sigma^{2} & = & \frac{1}{2}\taum\lambda\langle(w^{j})^{2}\rangle.\nonumber 
\end{eqnarray}
In the absence of a threshold, the stationary density follows from
the solution of $S(V,t)=0$ as a Gaussian with mean $\mu$ and variance
$\sigma^{2}$.

Equation \prettyref{eq:P(V,t)} and \prettyref{eq:infinitesimal_moments}
hold in general for excitatory events with i.i.d. random amplitudes
arriving at Poisson time points. Given the $K=cfN$ common excitatory
afferents' activities are generated by a MIP process, the number of
$k$ synchronized afferents follows a binomial distribution $P(k)=B(K,p,k)=\binom{K}{k}\, p^{k}(1-p)^{K-k}$,
with moments $\left\langle k\right\rangle =Kp$ and $\left\langle k^{2}\right\rangle =Kp(1-p+Kp)$.
The total rate $\frac{\nuin}{p}\langle k\rangle=\nuin K$ of arriving
events is independent of $p$, as is the contribution to the mean
membrane potential $\mu$. Further we assume the neurons to be contained
in a network that is in the balanced state, i.e. $g=f/(1-f)$, and
that all afferents have the same rate $\nuin$. Thus, excitation and
inhibition cancel in the mean so that $\mu=V_{0}$. Due to the independence
of excitatory and inhibitory spike trains they contribute additively
to the variance $\sigma^{2}$ in \prettyref{eq:infinitesimal_moments}.
The variance due to $(1-f)N$ inhibitory afferents with rate $\nuin$
is $(1-f)N\nuin g^{2}F_{2}$, with $F_{2}=\frac{1}{2}\tau_{m}w^{2}$.
An analog expression holds for the contribution of unsynchronized
disjoint excitatory afferents. The contribution of $K$ excitatory
afferents from the MIP follows from \eqref{eq:infinitesimal_moments}
as $\frac{\nuin}{p}\left\langle k^{2}\right\rangle F_{2}$. So together
we obtain

\begin{eqnarray}
\sigma^{2} & = & \left(cf(1-p+cfNp)+(1-c)f+g^{2}(1-f)\right)N\nuin F_{2}\label{eq:sigma2}\\
 & = & \left(f(1-cp+c^{2}fNp)+g^{2}(1-f)\right)N\nuin F_{2}.\nonumber 
\end{eqnarray}

\prettyref{fig:1}B shows that \prettyref{eq:sigma2} is in good agreement
with simulation results. We are further interested in describing the
correlation $\rho_{\inp}$ between the membrane potentials of both
neurons. The covariance is caused by the contribution from shared
excitation $\frac{\nuin}{p}\left\langle k^{2}\right\rangle F_{2}$,
in addition to the contribution from shared inhibition $c(1-f)N\nuin g^{2}F_{2}$,
which together result in the correlation coefficient

\begin{equation}
\rho_{\inp}=\left(f(1-p+cfNp)+g^{2}(1-f)\right)cN\nuin F_{2}/\sigma^{2}.\label{eq:rho_in}
\end{equation}

Again, \prettyref{fig:1}C shows that \prettyref{eq:rho_in} is in
good agreement with simulation results. In order to isolate and control
the effect of the synchrony parameter $p$ on \prettyref{eq:sigma2}
and \prettyref{eq:rho_in}, in the following we will contrast two
distinct scenarios: In the first scenario, we generate a certain amount
of input correlation $\rho_{\inp}$ using common input alone without
any spiking synchrony $p$. In the second scenario we generate the
same amount of input correlation $\rho_{\inp}$ by using a certain
amount of spike synchrony $p>0$. In order to have the same working
point in both cases, we need to keep the mean $\mu$ and variance
$\sigma^{2}$ of the single neuron's membrane potential constant for
a given fixed input correlation $\rho_{\inp}$. At $p=0$ the input
correlation $\rho_{\inp}$ \prettyref{eq:rho_in} equals the common
input fraction $c$ and does not depend on $\nuin$. For $p>0$ the
same input correlation can be achieved by an appropriate decrease
of $c$. To this end we solve \prettyref{eq:rho_in} for $c$, obtaining
the adjusted common input fraction $\bar{c}(\rho_{\inp},p)$ as the
positive root of the resulting quadratic equation. 

Since $\bar{c}(\rho_{\inp},p)$ does not depend on the afferent rate
we may adjust $\nuin$ in order to keep the total variance \prettyref{eq:sigma2}
on a constant level. At $p=0$ and firing rate $\nuin$ the variance
is given by $\sigma^{2}\bigr\rvert_{p=0}=\left(f+g^{2}(1-f)\right)N\nuin F_{2}$.
For $p>0$ the same variance can be achieved by an appropriate decrease
of $\nuin$. We therefore solve \prettyref{eq:sigma2} for $\nuin$,
obtaining the reduced firing rate $\bar{\nu}_{\mathrm{in}}(\sigma^{2},\rho_{\inp},p)$.

\begin{figure}
\begin{centering}
\includegraphics{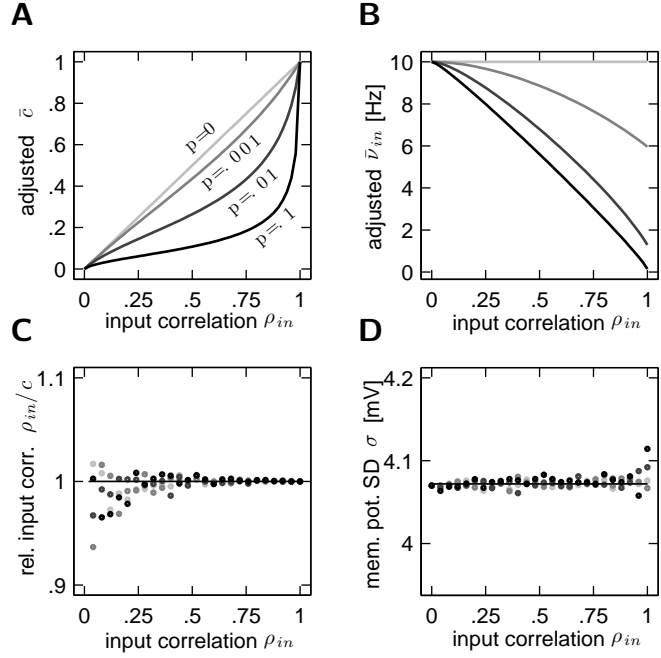}
\par\end{centering}

\caption{Isolation and control of the effect of synchrony on the free membrane
potential statistics. \textbf{A},\textbf{B} Adjusted common input
fraction $\bar{c}$ (\textbf{A}) and input firing rate $\bar{\nu}_{\mathrm{in}}$
(\textbf{B}) for different values of $p$ (gray coded) that ensure
the same variance and covariance as for $p=0$. \textbf{C} Correlation
coefficient $\rhoin$ normalized by $c$ between the free membrane
potential of a pair of neurons using the adjusted common input fraction
$\bar{c}$. \textbf{D} Standard deviation of the free membrane potential,
using the adjusted firing rate $\bar{\nu}_{\mathrm{in}}$. The statistics
of the free membrane potential measured in simulations in panels C
and D are further verified via \prettyref{eq:rho_in} and \prettyref{eq:sigma2}
(solid lines).\label{fig:2}}
\end{figure}

We evaluate this approach by simulating the free membrane potential
of a pair of leaky integrate-and-fire neurons driven by correlated
input. For different values of $p$, we choose $\bar{c}(\rho_{\inp},p)$
and $\bar{\nu}_{\mathrm{in}}(\sigma^{2},\rho_{\inp},p)$, shown in
\prettyref{fig:2}A and B, to keep the variance and the correlation
constant. \prettyref{fig:2}A shows that the adjustment of the common
input fraction becomes substantial only for higher values of $p$:
while for $p=0.001$ the reduced $\bc$ is only slightly smaller than
$c$, for $p=0.1$ and $\rhoin=0.8$ it is reduced to $\bar{c}=0.21$.
\prettyref{fig:2}B shows that even for small amounts of input synchrony,
$\nu_{\inp}$ needs to be decreased considerably in order to prevent
the increase of membrane potential variance (\prettyref{fig:1}B).
In the extreme case of $\rhoin=1$ and $p=0.1$ (both neurons receive
identical and strongly synchronous excitatory input) an initial input
firing rate of $10$ Hz needs to be decreased to $\bar{\nu}_{\mathrm{in}}=0.15$
Hz. \prettyref{fig:2}C and D confirm that indeed both the correlation
and the variance of the free membrane potential remain constant throughout
the whole range of $\rhoin$ and for all simulated values of $p$.

\subsection{Correlation Transmission\label{sub:Correlation-Transmission}}

\begin{figure}
\begin{centering}
\includegraphics{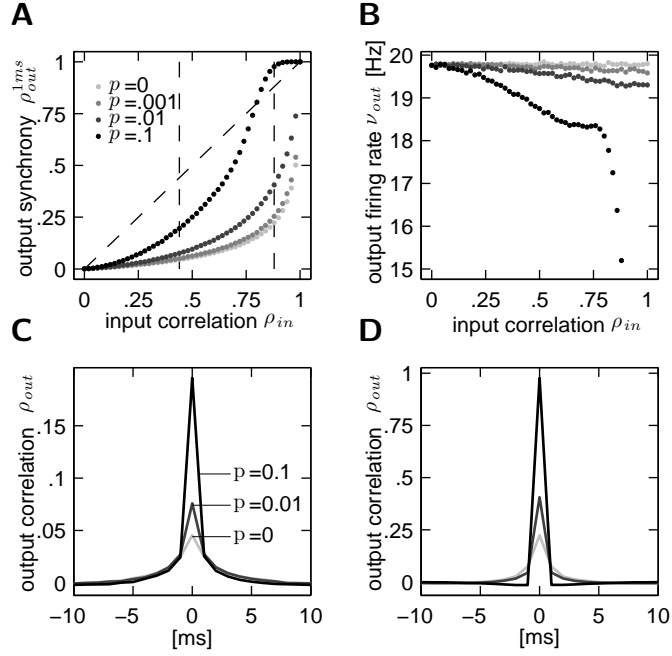}
\par\end{centering}

\caption{Correlation transmission of a pair of integrate-and-fire neurons.
\textbf{A} Output spike synchrony as a function of input correlation
$\rhoin$ and for four different values of input synchrony $p=0$,
$p=0.001$, $p=0.01$ and $p=0.1$ (gray-coded). Dashed black line
with slope $1$ indicates $\rhoout=\rhoin$. \textbf{B} Corresponding
mean output firing rate of the neurons.\textbf{ C},\textbf{D} Cross-correlation
functions at input correlations $\rhoin=0.44$ (C) and $\rhoin=0.88$
(D) (indicated by dashed vertical lines in A) for the three values
of input synchrony $p$ as indicated in C.\label{fig:3}}
\end{figure}

In order to study the transmission of correlation by a pair of neurons,
we need to ensure that the single neuron's working point does not
change with the correlation structure of the input. The diffusion
approximation \prettyref{eq:P(V,t)} suggests, that the decisive properties
of the marginal input statistics are characterized by the first ($\mu$)
and second moment ($\sigma^{2}$). As we supply balanced spiking activity
to each neuron, the mean $\mu$ is solely controlled by the resting
potential $V_{0}$, as outlined above. For any given value of $p$
and $\rhoin$, choosing the afferent rate $\bar{\nu}_{\mathrm{in}}(\sigma^{2},\rhoin,p)$
ensures a constant second moment $\sigma^{2}$. Consequently, \prettyref{fig:3}B
confirms that the fixed working point ($\mu,\sigma^{2}$) results
in an approximately constant neural firing rate $\nuout$ for weak
to moderate input synchrony $p$. For strong synchrony, fluctuations
of the membrane potential become non-Gaussian and the firing rate
decreases; the diffusion approximation breaks down. 

In studies which investigate the effect of common input on the correlation
transmission of neurons, the input correlation is identical to the
common input fraction $c$ \cite{DeLaRocha07_802,Shea-Brown08}. In
the presence of input synchrony this is obviously not the case (\prettyref{fig:1}C).
Choosing the afferent rate and the common input fraction according
to $\bar{\nu}_{\mathrm{in}}(\sigma^{2},\rhoin,p)$ and $\bar{c}(\rho_{\inp},p)$,
respectively, enables us to realize the same input correlation $\rhoin$
with different contributions from shared inputs and synchronized events.
We may thus investigate how the transmission of correlation by a neuron
pair depends on the relative contribution of synchrony to the input
correlation $\rhoin$. \prettyref{fig:3}A shows the output synchrony
as a function of $\rhoin$ for four fixed values of input synchrony
$p$. As the input correlation is by construction the same for all
values of $p$, changes in the output synchrony directly correspond
to a different correlation transmission coefficient. Even weak spiking
synchrony ($p=0.001$) in the common input effectively increases the
output synchrony, compared to the case where the same input correlation
is exclusively caused by common input ($p=0$). Stronger synchrony
($p=0.01$ and $p=0.1$) further increases this transmission. In \prettyref{fig:3}B
we confirm that the increase of output spike synchrony is not caused
by an increase of the output firing rate of the neurons, but rather
their rate remains constant up to intermediate values of $p\le0.01$.
The drastic decrease of the output firing rate for $p=0.1$ does not
rebut our point, but rather strengthens it: correlation transmission
is expected to decrease with lower firing rate \cite{sheabrown07,Shea-Brown08}
for Gaussian inputs. However, here we observe the opposite effect
in the case of strongly non-Gaussian inputs due to synchronous afferent
spiking. We will discuss this issue in the following paragraph, deriving
an analytical prediction for the correlation transmission. Moreover,
we observe that the increased transmission is accompanied by a sharpening
of the correlation function with respect to the case of $p=0$ (cf.
\prettyref{fig:3}C and D).

For correlated inputs caused by common inputs alone (no synchrony,
$p=0$) or by weak spiking synchrony ($p\le0.01$) the transmission
curves in \prettyref{fig:3}A are always below the identity line.
This means that the neural dynamics does not transmit the correlation
perfectly, but rather causes a decorrelation. Recent work has shown
that the finite life time of the memory stored in the membrane voltage
of a leaky integrate-and-fire neuron is directly related to this decorrelation
\cite{Rosenbaum10_00116}. Quantitative approximations of this decorrelation
by non-linear threshold units can be understood in the Gaussian white
noise limit \cite{DeLaRocha07_802,Shea-Brown08}. For input correlation
caused by spiking synchrony, however, we observe a qualitatively new
feature here. In the presence of strong spiking synchrony ($p=0.1$),
in the regime of high input correlation ($\rhoin\gtrsim0.8$) the
correlation transmission coefficient exceeds unity. In other words,
the neurons correlate their spiking activity at a level that is higher
than the correlation between their inputs. In order to obtain a quantitative
understanding of this boost of correlation transmission by synchrony,
in the following two subsections we will in turn investigate the mechanisms
in the limit of low and high input correlations, respectively.

\subsection{Correlation Transmission in the Low Correlation Limit\label{sub:low_corr_limit}}

For Gaussian white noise input and in the the limit of low input correlation,
the correlation transmission can be understood by standard methods
\cite{Morenobote06_028101,MorenoBote08,DeLaRocha07_802,Shea-Brown08}.
This diffusion approximation assumes that the amplitudes of synaptic
events are infinitesimally small. For uncorrelated Poisson processes
and large number of afferents $N$, the theory is still a fairly good
approximation. For small synaptic jumps approximate extensions are
known \cite{helias10,Helias10_1000929} and exact results can be obtained
for jumps with exponentially distributed amplitudes \cite{Richardson10_178102}.
However, in order to treat spiking synchrony in the common input to
a pair of neurons, we need to extend the perturbative approach here.

\begin{figure}
\begin{centering}
\includegraphics{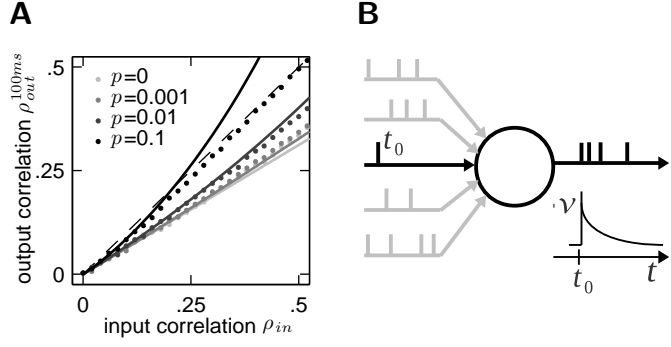}
\par\end{centering}

\caption{Approximation of the output correlation in the limit of low input
correlation.\textbf{ A} Correlation transmission in the low input
correlation limit. Data points show the output correlation $\rho_{\out}^{100\ms}$
resulting from simulations, solid lines show the theoretical approximation
$\rho_{\out}^{\infty}$ \prettyref{eq:integral_correlation}. Dashed
black line indicates $\rhoout=\rhoin$. \textbf{B} Deflection of the
firing rate with respect to base rate caused by an additional synaptic
event at $t=t_{0}$.\label{fig:4}}
\end{figure}

Before deriving an expression for the correlation transmission by
a pair of neurons, we first describe the firing rate deflection of
a neuron caused by a single additional synaptic impulse of amplitude
$J$ at $t=0$ on top of synaptic background noise. Within the diffusion
approximation, the total afferent input to the neuron can be described
by the first two moments $\mu$ and $\sigma^{2}$ \prettyref{eq:infinitesimal_moments}.
We determine the integral over the excursion of the firing rate $h(t,J)=\langle s(t|\,\textrm{impulse of amplitude}\, J\,\textrm{at}\, t=0)-\nuout\rangle$
caused by the additional impulse with respect to the base rate $\nuout$,
as illustrated in \prettyref{fig:4}B. An additional input with a
stationary Poisson rate $\nu$ has an effect $\tau_{m}J\nu$ on the
mean and $\frac{1}{2}\tau_{m}J^{2}\nu$ on the variance. The integral
of the impulse response can be expressed as $\Pint(J)\equiv\int_{0}^{\infty}h(t,J)\; dt=\left.\frac{\partial\nuout(\nu)}{\partial\nu}\right|_{\nu=0}$
\cite{Helias10_1000929}, where $\nuout(\nu)$ is the equilibrium
firing rate of the neuron receiving synaptic input causing a mean
$\mu+\tau_{m}J\nu$ and a variance $\sigma^{2}+\frac{1}{2}\tau_{m}J^{2}\nu$.
Using the well known expression for the mean first passage time \cite{Siegert51,Brunel00}
\begin{eqnarray}
\nu_{\out}^{-1}(\mu,\sigma) & = & \tau_{r}+\sqrt{\pi}\tau_{m}\left(F(y_{\theta})-F(y_{r})\right)\label{eq:nu_Siegert}\\
\mathrm{with}\nonumber \\
F(y) & = & \int^{y}f(y)\; dy\qquad f(y)=e^{y^{2}}(\mathrm{erf}(y)+1)\nonumber \\
y_{\theta} & = & \frac{V_{\theta}-\mu}{\sqrt{2}\sigma}\qquad y_{r}=\frac{V_{r}-\mu}{\sqrt{2}\sigma}\nonumber 
\end{eqnarray}
and applying the chain rule we arrive at

\begin{eqnarray}
\Pint(J) & = & \int_{0}^{\infty}h(t,J)\; dt\label{eq:P_int}\\
\textrm{} & = & \alpha J+\beta J^{2}\nonumber \\
\textrm{with}\nonumber \\
\alpha & = & (\nuout\tau_{m})^{2}\sqrt{\frac{\pi}{2}}\frac{1}{\sigma}\left(f(y_{\theta})-f(y_{r})\right)\nonumber \\
\beta & = & (\nuout\tau_{m})^{2}\sqrt{\pi}\frac{1}{4\sigma^{2}}\left(f(y_{\theta})y_{\theta}-f(y_{r})y_{r}\right),\nonumber 
\end{eqnarray}
where $\Pint$ is the expected number of additional spikes over baseline
caused by the injected pulse. Assuming that the incoming events are
pairwise correlated by a MIP \cite{Kuhn03_67} so that the sum of
synchronously arriving events $k$ is binomially distributed, we can
expand \prettyref{eq:P_int} and set $J=kw$, with $k\sim B(\bar{c}fN,p,k)$
to obtain

\begin{eqnarray*}
\Pint & = & w\alpha\sum_{k}k\cdot B(\bar{c}fN,p,k)+w^{2}\beta\sum_{k}k^{2}\cdot B(\bar{c}fN,p,k)\\
 & = & w\alpha M_{1}+w^{2}\beta M_{2},
\end{eqnarray*}
with $M_{1}$ and $M_{2}$ being the first and second moments of the
binomial distribution%
\footnote{The first four moments of the binomial distribution $B(N,p)$ are

$M_{1}=Np$,

$M_{2}=Np(1-p+Np)$,

$M_{3}=Np(1-3p+3Np+2p^{2}-3Np^{2}+N^{2}p^{2})$ and

$M_{4}=Np(1-7p+7Np+12p^{2}-18Np^{2}+6N^{2}p^{2}-6p^{3}+11Np^{3}-6N^{2}p^{3}+N^{3}p^{3})$
\label{fn:moments}%
}.

If two neurons receive statistically the same input they fire with
the same firing rate. Furthermore, due to the common input they both
receive the same synchronous events with binomially distributed amplitudes
$J=kw$. Any such event causes the same amount of excess spikes $\Pint(kw)$
in both neurons. If $s_{i}(t)$ denotes the spike train of neuron
$i=1,2$, the cross covariance function is defined as

\begin{equation}
\kappa_{\mathrm{out}}(\Delta)=\langle(s_{1}(\Delta+t)-\nuout)(s_{2}(t)-\nuout)\rangle_{t},\label{eq:cross_covariance_function}
\end{equation}
where the expectation value $\langle\rangle_{t}$ is taken over realizations
of the input and over time $t$. $\kappa_{\mathrm{out}}(\Delta)$
drops to zero for $\Delta\rightarrow\infty$. The only cause of covariation
of the firing probability are the common events in the input. The
sequence of events of amplitude $J$ arriving with rate $\lambda_{J}$
contribute to the integrated covariance function as

\begin{eqnarray*}
\bar{\kappa}_{\mathrm{out}}(J) & = & \intop_{-\infty}^{\infty}\kappa_{\out}(\Delta,J)\; d\Delta\\
 & = & \intop_{-\infty}^{\infty}\lim_{T\rightarrow\infty}(2T\lambda_{J})\frac{1}{2T}\intop_{-T}^{T}(s_{1}(\Delta+t)-\nuout)(s_{2}(t)-\nuout)\, dt\, d\Delta\\
 & = & \lambda_{J}\intop_{-\infty}^{\infty}(s_{1}(t)-\nuout)\, dt\quad\intop_{-\infty}^{\infty}(s_{2}(t)-\nuout)\, dt\\
 & = & \lambda_{J}\intop_{0}^{\infty}h(t,J)\; dt\;\intop_{0}^{\infty}h(t,J)\; dt\\
 & = & \lambda_{J}\Pint(J)^{2}.
\end{eqnarray*}
The second step holds by exchanging the order of integrals and by
inserting the expected number of events $2T\lambda_{J}$ during time
$2T$. We now need to treat the excitatory and inhibitory input differently,
since inhibitory events are not synchronized and furthermore additionally
weighted by $-g$. Averaging over all common input events, the effective
rate of an event of amplitude $J=kw$ is given by $\lambda_{kw}=\lambda_{m}B(\bar{c}fN,p,k)$
we therefore obtain

\[
\bar{\kappa}_{\mathrm{out}}=\lambda_{m}\sum_{k}B(\bar{c}fN,p,k)\cdot\Pint(kw)^{2}+\bar{\nu}\bar{c}(1-f)N\cdot\Pint(-gw)^{2},
\]
 and use $\Pint^{2}(J)=\alpha^{2}J^{2}+2\alpha\beta J^{3}+\beta^{2}J^{4}$
to obtain

\begin{eqnarray}
\bar{\kappa}_{\mathrm{out}} & = & \lambda_{m}\left(\alpha^{2}w^{2}M_{2}+2\alpha\beta w^{3}M_{3}+\beta^{2}w^{4}M_{4}\right)\label{eq:integral_covariance}\\
 &  & +\bar{\nu}\bar{c}(1-f)N\left(\alpha^{2}(-gw)^{2}+2\alpha\beta(-gw)^{3}+\beta^{2}(-gw)^{4}\right),\nonumber 
\end{eqnarray}
 where $M_{3}$ and $M_{4}$ are the third and fourth moment of the
binomial distribution$\footref{fn:moments}$.

In the limit of low input correlation we are interested in the output
correlation on long time scales. In order to obtain this quantity,
we need to normalize the integral of \prettyref{eq:integral_covariance}
by the integral of the auto-covariance of the neurons' spike trains.
This integral equals $\mathrm{FF}\nuout$ \cite{MorenoBote08}, with
the Fano factor $\mathrm{FF}$. In the long time limit the Fano factor
of a renewal process equals the squared coefficient of variation $CV^{2}$
\cite{Cox62}, which can be calculated in the diffusion limit \cite[App. A1]{Brunel00}.
Thus, we obtain 
\begin{equation}
\rho_{\out}^{\infty}\simeq\frac{\bar{\kappa}_{\out}}{CV^{2}\nuout}.\label{eq:integral_correlation}
\end{equation}

\prettyref{fig:4}A shows that the output spike correlation of a pair
of neurons is fairly well approximated by $\rho_{\out}^{\infty}$
in the lower correlation regime. While the approximation is good over
almost the whole displayed range of $\rho_{\inp}$ for $p=0.001$
and $p=0.01$, for $p=0.1$ the theory only works for values of $\rho_{in}<0.3$
in agreement with previous studies \cite{sheabrown07,Shea-Brown08}
applying a similar perturbative approach to the case of Gaussian input
fluctuations.

\subsection{Correlation Transmission in the High Correlation Limit\label{sub:high_corr_limit}}

\begin{figure}
\begin{centering}
\includegraphics{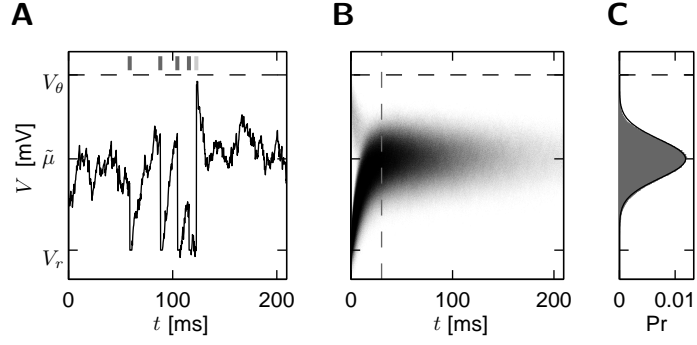}
\par\end{centering}

\caption{Neural dynamics in the regime of high input correlation and strong
synchrony.\textbf{ A} Exemplary time course of a membrane potential
driven by input containing strong, synchronous spike events. During
the time period shown, five MIP events arrive (indicated by tick marks
above $V_{\theta}$). The first four drive the membrane potential
above the threshold $V_{\theta}$, after which $V$ is reset to $V_{r}$
and the neuron emits a spike (dark gray tick marks above $V_{\theta}$).
The fifth event is not able to deflect $V$ above threshold (light
gray) and the membrane potential quickly repolarizes towards its steady
state mean $\tilde{\mu}$ (see text). \textbf{B} Time-resolved membrane
potential probability density $P(V,t)$ triggered on the occurrence
of a MIP event at $t=0$. Since most MIP events elicit a spike, after
resetting $V$ to $V_{r}$ the membrane potential quickly depolarizes
and settles to a steady state Gaussian distribution. The slight shade
of gray observable for small $t$ just below the threshold $V_{\theta}$
is caused by the small amount of MIP events that were not able to
drive the membrane potential above threshold. \textbf{C} Probability
density of the membrane potential in steady state. Theoretical approximation
(black) was computed using $\tilde{\mu}$ and $\tilde{\sigma}$ (see
text and \prettyref{eq:Vt}), empirical measurement (gray) was performed
for $t>30$ ms (gray dashed line in B). Simulation parameters were
$p=0.1$, $\bar{c}=0.26$ ($c=0.87$) and $\bar{\nu}=1.75$ ($\nuin=10$)
Hz. Other parameters as in \prettyref{tab:sim_parameters-1}.\label{fig:5}}
\end{figure}

In order to understand how the neurons are able to achieve a correlation
coefficient larger than one, we need to take a closer look at the
neural dynamics in the high correlation regime. We refer to the strong
pulses caused by synchronous firing of numerous afferents as MIP events.
\prettyref{fig:5}A shows an example of the membrane potential time
course that is driven by input in the high correlation regime. At
sufficiently high synchrony as shown here, most MIP events elicit
a spike in the neuron, whereas fluctuations due to the disjoint input
alone are not able to drive the membrane potential above threshold.
Thus, in between two MIP events the membrane potential distribution
of each neuron evolves independently and fluctuations are caused by
the disjoint input alone. \prettyref{fig:5}B shows the time-dependent
probability density of the membrane potential, triggered on the time
of arrival of a MIP event. We observe that most MIP events cause an
action potential, followed by the recharging of the membrane after
it has been reset to $V_{r}$ at $t=0$. After a short period of repolarization
the membrane potential quickly reaches its steady state. The contribution
of the $\bar{c}fN$ common, excitatory afferents to the membrane potential
statistics is limited to those occasional strong depolarizations.
Between two such events they neither contribute to the mean nor to
the variance of $V$. Hence the effective mean and variance of the
membrane potential are due to the disjoint input alone,\textrm{ given
by $\tilde{\mu}=V_{0}-g(1-f)\bar{c}N\bar{\nu}_{\mathrm{in}}F_{1}$
and $\tilde{\sigma}^{2}=\left[f(1-\bar{c})+g^{2}(1-f)\right]N\bar{\nu}_{\mathrm{in}}F_{2}$
with $F_{1}=\tau_{m}w$ and $F_{2}=\frac{1}{2}\tau_{m}w\text{\texttwosuperior}$.}
\prettyref{fig:5}C shows in gray the empirical distribution of the
membrane potential between two MIP events after it has reached the
steady state. It is well approximated by a Gaussian distribution with
mean $\tilde{\mu}$ and variance $\tilde{\sigma}$. The membrane potential
can therefore be approximated as a threshold-free Ornstein-Uhlenbeck
process \cite{Uhlenbeck30,Tuckwell88a}.

Let us now recapitulate these last thoughts in terms of a pair of
neurons: In the regime of synchronized high input correlation (e.g.
$p=0.1$, $\rhoin>0.8$), MIP events become strong enough so that
most of them elicit a spike in both neurons. At the same time, the
uncorrelated, disjoint sources (which can be considered as sources
of noise) induce fluctuations of the membrane potential which are,
however, not big enough to drive the membrane potential above threshold.
Thus, while the input to both neurons still contains a considerable
amount of independent noise, their output spike trains are (for sufficiently
high $\rhoin$) a perfect duplicate of the mother spike train that
generates the MIP events in their common excitatory input, explaining
the observed correlation transmission coefficient larger than unity.
Note that this is the reason for the drastic decrease of the output
firing rate in \prettyref{fig:3}B, which in the limit of high input
correlation approaches the adjusted input firing rate $\bar{\nu}_{\mathrm{in}}$
(\prettyref{fig:2}B).

We would like to obtain a qualitative assessment of the correlation
transmission in the high correlation input regime. Since the probability
of output spikes caused by the disjoint sources is vanishing, the
firing due to MIP events inherits the Poisson statistics of the mother
process. Consequently, the autocovariance function of each neurons'
output spike train is a $\delta$-function weighted by its rate $\nu_{0}=\lambda_{m}P_{\mathrm{inst}}$,
where $P_{\mathrm{inst}}$ is the probability that a MIP event triggers
an outgoing spike in one of the neurons. The output correlation can
hence be approximated by the ratio

\begin{equation}
\rho_{\mathrm{out}}\simeq\frac{P_{\mathrm{sync}}}{P_{\mathrm{inst}}},\label{eq:rho_out}
\end{equation}
where $P_{\mathrm{sync}}$ is the probability that a MIP event triggers
an outgoing spike in both neurons at the same time. Note that the
approximation \eqref{eq:rho_out} holds for arbitrary time scales,
as the spike trains have Poisson statistics in this regime. In order
to evaluate $P_{\mathrm{inst}}$ and $P_{\mathrm{sync}}$, we use
the simplifying assumption that the last MIP event at $t=0$ caused
a reset of the neuron to $V_{r}=0$, so the distribution $P(V,t)$
of the membrane potential evolves like an Ornstein-Uhlenbeck process
as \cite{Tuckwell88a}
\begin{align}
P(V,t)= & \frac{1}{\sqrt{2\pi\tilde{\sigma}(t)^{2}}}\exp\left(-{\displaystyle \frac{(V-\tilde{\mu}(t))^{2}}{2\tilde{\sigma}(t)^{2}}}\right)\label{eq:Vt}\\
\text{with}\nonumber \\
\tilde{\mu}(t)= & \tilde{\mu}\left(1-e^{-\frac{t}{\tau_{m}}}\right)\nonumber \\
\tilde{\sigma}^{2}(t)= & \tilde{\sigma}^{2}\left(1-e^{-\frac{2t}{\tau_{m}}}\right),\nonumber 
\end{align}
 which is the solution of \prettyref{eq:P(V,t)} with initial condition
$V(0)=0$. We evaluate $P_{\mathrm{inst}}$ from the probability mass
of the voltage density shifted across threshold by an incoming MIP
event as

\begin{equation}
P_{\mathrm{inst}}=\sum_{k=1}^{\bar{c}fN}B(\bar{c}fN,p,k)\cdot\intop_{0}^{\infty}dt\;\lambda_{m}S(t)\cdot\intop_{V_{\theta}-kw}^{V_{\theta}}dV\; P(V,t),\label{eq:P_inst_old}
\end{equation}
where the survivor function $S(t)=\exp(-\lambda_{m}t)$ is the probability
that after a MIP event occurred at $t=0$ the next one has not yet
occurred at $t>0$. So $dt\,\lambda_{m}S(t)$ is the probability that
no MIP event has occurred in $\left[0,t\right]$ and it will occur
in $\left[t,t+dt\right]$ \cite{Cox62}. The binomial factor $B$
is the probability for the amplitude of a MIP event to be $kw$ and
the last integral is the probability that a MIP event of amplitude
$kw$ causes an output spike \cite{Helias10_1000929}. We first express
$I(V,t)=\int_{V}^{V_{\theta}}dV\; P(V,t)$ in terms of the error function
using \prettyref{eq:Vt} with the substitution $x=\frac{V-\tilde{\mu}(t)}{\sqrt{2}\tilde{\sigma}(t)}$,
to obtain

\begin{eqnarray}
I(V,t) & = & \frac{1}{2}\left[\mathrm{erf\left(\frac{V_{\theta}-\tilde{\mu}(t)}{\sqrt{2}\tilde{\sigma}(t)}\right)}-\mathrm{erf\left(\frac{V-\tilde{\mu}(t)}{\sqrt{2}\tilde{\sigma}(t)}\right)}\right],\label{eq:P_threshold}
\end{eqnarray}
where we used the definition of the error function $\mathrm{erf}(x)=\frac{2}{\sqrt{\pi}}\int_{0}^{x}e^{-x^{2}}\: dx$.
We further simplify the first integral in \prettyref{eq:P_inst_old}
with the substitution $y=e^{-\lambda_{m}t}$ to

\begin{eqnarray*}
\intop_{0}^{\infty}dt\: S(t)\cdot I(V,t) & = & -\intop_{1}^{0}\frac{dy}{\lambda y}y\cdot I\left(V,\frac{ln\, y}{-\lambda_{m}}\right),
\end{eqnarray*}
thus finally obtaining

\begin{equation}
P_{\mathrm{inst}}=\sum_{k=1}^{\bar{c}fN}B(\bar{c}fN,p,k)\cdot\intop_{0}^{1}\hat{I}(V_{\theta}-kw,y)\; dy,\label{eq:P_inst}
\end{equation}
where we introduced $\hat{I}(V,y)$ as a shorthand for \prettyref{eq:P_threshold}
with $\tilde{\mu}(t)$ and $\tilde{\sigma}(t)$ expressed in terms
of the substitution variable $y$ as $\hat{\mu}(y)=\tilde{\mu}\left(1-y^{\frac{1}{\lambda_{m}\tau_{m}}}\right)$
and $\hat{\sigma}(y)=\tilde{\sigma}\left(1-y^{\frac{2}{\lambda_{m}\tau_{m}}}\right)$,
following from \prettyref{eq:Vt}. In order to approximate the probability
$P_{\mathrm{sync}}$ that the MIP event triggers a spike in both neurons
we need to square the second integral in \prettyref{eq:P_inst_old},
because the voltages driven by disjoint input alone are independent,
so their joint probability distribution factorizes, leading to

\begin{equation}
P_{\mathrm{sync}}=\sum_{k=1}^{\bar{c}fN}B(\bar{c}fN,p,k)\cdot\intop_{0}^{1}\hat{I}(V_{\theta}-kw,y)^{2}\, dy.\label{eq:P_sync}
\end{equation}

\begin{figure}
\begin{centering}
\includegraphics{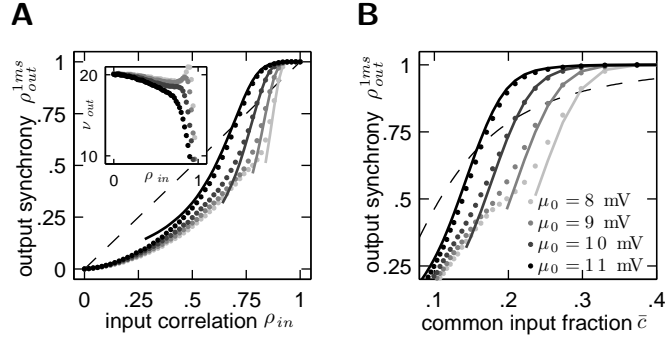}
\par\end{centering}

\caption{Approximation of the output synchrony in the limit of high input correlation.\textbf{
A} Output spike synchrony as a function of input correlation in the
limit of high input correlation and strong synchrony $p=0.1$. Data
points and solid lines show results from simulations and theoretical
approximation \prettyref{eq:rho_out}, respectively. Gray code corresponds
to the four different mean membrane potential values $\mu_{0}$ as
depicted in B, the input firing rate $\nuin$ was $17.5\Hz$, $13.7\Hz$,
$10.0\Hz$ and $7.2\Hz$, correspondingly (from low to high $\mu_{0}$).
The working point used in the previous sections corresponds to $\mu_{0}=10\mV$,
$\nuin=10\Hz$. The inset shows the output firing rate at the four
working points.\textbf{ B} Output spike synchrony as a function of
the actual common input fraction $\bar{c}$ at the four working points.
Dashed curves in A and B indicate $\rhoout=\rhoin$.\label{fig:6}}
\end{figure}

Having defined $P_{\mathrm{inst}}$ and $P_{\mathrm{sync}}$ we now
can calculate $\rho_{\mathrm{out}}$ using \prettyref{eq:rho_out}
in the high input correlation regime. So far, we have considered both
neurons operating at a fixed working point, defined by the mean and
variance \prettyref{eq:infinitesimal_moments}. Due to the non-linearity
of the neurons we expect the effect of synchronous input events on
their firing to depend on the choice of this working point. We therefore
performed simulations and computed \prettyref{eq:spike_count_correlation}
using four different values for the mean membrane potential $\mu_{0}$
(\prettyref{fig:6}). This was achieved by an appropriate choice of
a DC input current $I_{0}$ and accordingly adjusting the input firing
rate $\nuin$ in order to keep the mean firing rate constant (\prettyref{fig:6}A,
inset). The data points from simulations in \prettyref{fig:6}A show
that different working points of the neurons considerably alter the
correlation transmission in the limit of high input correlation. At
working points near the threshold ($\mu_{0}=11\mV$) MIP events more
easily lead to output spikes, thereby boosting the transmission of
correlation, as compared to working points that are further away from
the threshold ($\mu_{0}=8\mV$). Solid lines in \prettyref{fig:6}A
furthermore show that \prettyref{eq:rho_out} indeed provides a good
approximation of the output spike correlation when the input to both
neurons is strongly synchronized. Obviously, the assumption has to
hold that the probability density of the membrane potential is sufficiently
far from the threshold, which for $p=0.1$ is only the case if $\rhoin\gtrsim0.75$.
Hence, the approximation becomes less accurate for lower input correlations,
as expected. Note that, as opposed to \prettyref{fig:1}E, the effective
common input fraction $\bar{c}$ in \prettyref{fig:6}A is much lower
than $\rhoin$. \prettyref{fig:6}B shows the same data as a function
of the actual fraction of shared afferents $\bar{c}$. It reveals
that the gain of correlation transmission above unity is already reached
at fractions of common input as low as $c=0.15$ (for $\mu_{0}=11\mV$),
which is a physiologically plausible value.

\section{Discussion\label{sec:Discussion}}

\subsection{Summary of Results}

In this work we investigate the correlation transmission by a neuron
pair, using two different types of input spike correlations. One is
caused solely by shared input -- typically modeled as Gaussian white
noise in previous studies \cite{DeLaRocha07_802,Shea-Brown08} --
while in the other the spikes in the shared input may additionally
arrive in synchrony. In order to shed light on the question whether
cortical neurons operate as integrators or as coincidence detectors
\cite{Abeles82b,Koenig96,Hong12_1413}, we investigate their efficiency
in detecting and transmitting spike correlations of either type. We
showed that the presence of spike synchrony results in a substantial
increase of correlation transmission, suggesting that synchrony is
a prerequisite in explaining the experimentally observed excess spike
synchrony \cite{Riehle97_1950,Maldonado08_1523,Kilavik09_12653},
rather than being an epiphenomenon of firing rate due to common input
given by convergent connectivity \cite{Shadlen98}.

To model correlated spiking activity among the excitatory afferents
in the input to a pair of neurons we employ the Multiple Interaction
Process (MIP) \cite{Kuhn03_67}, resulting in non-Gaussian fluctuations
in the membrane potential of the receiving neurons. In this model
the parameter $p$ defines the pairwise correlation coefficient between
each pair of $N$ spike trains. If $N$ is large enough and all spike
trains are drawn independently ($p=0$) the summation of all $N$
spike trains is approximately equivalent to a Gaussian white noise
process \cite{Risken96,Tuckwell88a}. However, introducing spike correlations
between the spike trains ($p>0$) additionally allows for the modeling
of non-Gaussian fluctuating inputs. Such correlations have a strong
effect on the membrane potential statistics and the firing characteristics
of the neurons. The fraction of common input $c$ and the synchrony
strength $p$ each contribute to the total correlation between the
inputs to both neurons. We show how to isolate and control the effect
of input synchrony such that (1) a particular input correlation $\rhoin$
can be realized by an (almost) arbitrary combination of input synchrony
$p$ and common input fraction $c$, and (2) the output firing rate
of the neurons does not increase with $p$. This enables a fair comparison
of transmission of correlation due to input synchrony and due to common
input. We find that the non-linearity of the neuron model boosts the
correlation transmission due to the strong fluctuations caused by
the common source of synchronous events.

Given a fixed input correlation, the correlation transmission increases
with $p$. Most notably, this is the case although the output firing
rate of the neurons does not increase and is for the most part constant,
suggesting that the correlation susceptibility of neurons is not a
function of rate alone, as previously suggested \cite{DeLaRocha07_802},
but clearly depends on pairwise synchrony in the input ensemble. Previous
studies have shown that the transmission of correlation of neuron
pairs driven by Gaussian white noise can be approximated by employing
Fokker-Planck theory and perturbation theory \cite{Morenobote06_028101,MorenoBote08,DeLaRocha07_802,Shea-Brown08}.
In order to understand the effect of synchrony on the correlation
transmission we extended this approximation to synaptic input of finite
amplitudes. In the limit of low input correlation this extension indeed
provides a good approximation of the output correlation caused by
inputs containing spike correlations. Alternative models that provide
analytical results are those of thresholded Gaussian models \cite{Burak09_2269,Tchumatchenko10_058102}
or random walk models \cite{Rosenbaum10_00116}. In order to study
transmission in networks with different architecture than the simple
feed-forward models employed here, our results may be extended by
techniques to study simple network motifs developed in \cite{Ostojic09_10234}.

Hitherto existing studies argue that neurons either loose correlation
when they are in the fluctuation driven regime or at most are able
to preserve the input correlation in the mean driven regime \cite{Rosenbaum11_1261}.
Here, we provide evidence for a qualitatively new mechanism which
allows neurons to exhibit more output correlation than they receive
in their input. \prettyref{fig:3}A shows that in the regime of high
input correlation the correlation transmission coefficient can exceed
unity. This effect is observed at realistic values of pairwise correlations
($p\simeq0.1$) and common input fractions ($c\simeq0.25$). We provide
a quantitative explanation of the mechanism that enables neurons to
exhibit this behavior. We show that in this regime of high input correlation
$\rhoin$ the disjoint sources and the common inhibitory sources do
not contribute to the firing of the neurons, but rather the neurons
only fire due to the strong synchronous events in the common excitatory
afferents. Based on this observation, in this contribution we derive
an analytic approximation of the resulting output correlation beyond
linear perturbation theory that is in good agreement with simulation
results.

\subsection{Mechanism of Noise Suppression by Coincidence Detection}

\begin{figure}
\begin{centering}
\includegraphics[scale=0.8]{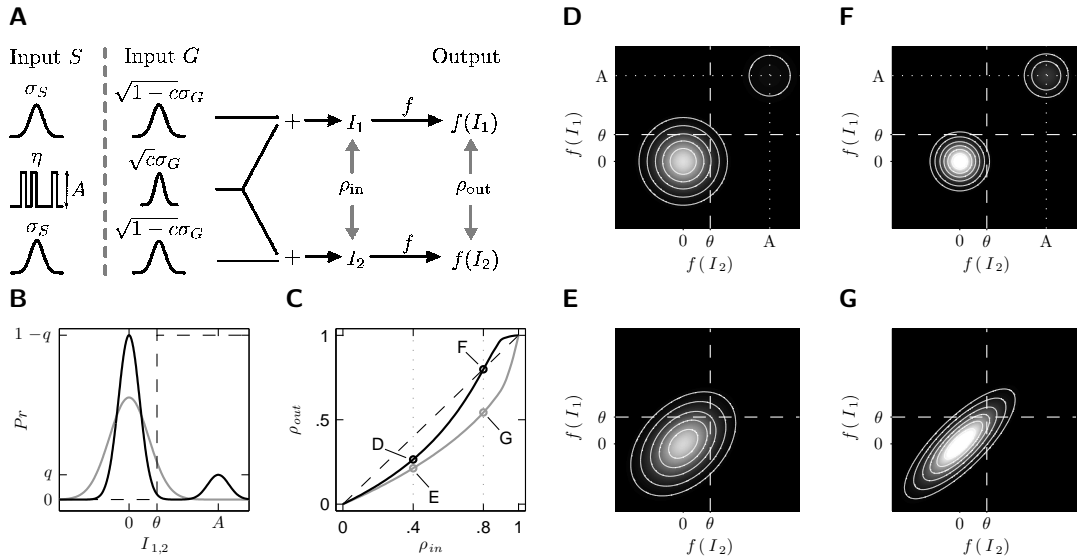}
\par\end{centering}

\centering{}\caption{Mechanistic model of enhanced correlation transmission by synchronous
input events. \textbf{A} The detailed model discussed in the results
section is simplified two-fold. 1) We consider binary neurons with
a static non-linearity $f(x)=H(x-\theta)$. 2) We distinguish two
representative scenarios with different models for the common input:
$G$: Gaussian white noise with variance $c\sigma_{G}^{2}$, representing
the case without synchrony, or $S$: a binary stochastic process $\eta(t)$
with constant amplitude $A$, mimicking the synchronous arrival of
synaptic events. In both scenarios in addition each neuron receives
independent Gaussian input. \textbf{B} Marginal distribution of the
total input $I_{1,2}$ to a single neuron for input $G$ (gray) and
$S$ (black) and for $\rhoin=0.8$. In input $S$ the binary process
$\eta$ alternates between $0$ (with probability $1-q$) and $A$
(with probability $q$), resulting in a bimodal marginal distribution.
The mean activity of one single neuron is given by the probability
mass above threshold $\theta$. We choose the variances $\sigma_{G}^{2}$
and $\sigma_{S}^{2}$ of the disjoint Gaussian fluctuating input such
that the mean activity is the same in both scenarios. \textbf{C} Output
correlation $\rho_{\mathrm{out}}=\frac{\mathrm{Cov}[f(I_{1}),f(I_{2})]}{\sqrt{\mathrm{Var}[f(I_{1})]\mathrm{Var}[f(I_{2})]}}$
as a function of the input correlation $\rho_{\mathrm{in}}$ (see
A) between the total inputs $I_{1}$ and $I_{2}$. Probability $q$
is chosen such that inputs $G$ and $S$ result in the same input
correlation $\rho_{\mathrm{in}}$. The four markers correspond to
the panels D - G. \textbf{D}-\textbf{G} Joint probability density
of the inputs $I_{1}$, $I_{2}$ to both neurons. For two different
values of $\rho_{\mathrm{in}}$ the lower row (E,G) shows the scenario
$G$, the upper row (D,F) the scenario $S$. Note that panel B is
the projection of the joint densities in F and G to one axis. Brighter
gray levels indicate higher probability density, same gray scale for
all four panels.\label{fig:7}}
\end{figure}

We presented a quantitative description of the increased correlation
transmission by synchronous input events for the leaky integrate-and-fire
(LIF) model. Our analytical results explain earlier observations from
a simulation study modeling synchrony by co-activation of a fixed
fraction of the excitatory afferents \cite{Stroeve01}. However, the
question remains what the essential features are that cause this effect.
An even simpler model consisting of a pair of binary neurons is sufficient
to qualitatively reproduce our findings and to demonstrate the generality
of the phenomenon for non-linear units, allowing us to obtain a mechanistic
understanding. In this model, whenever the summed input $I_{1,2}$
exceeds the threshold $\theta$ the corresponding neuron is active
($1$) otherwise it is inactive ($0$). In \prettyref{fig:7} we consider
two different implementations of input correlation, one using solely
Gaussian fluctuating common input (input $G$), the other representing
afferent synchrony by a binary input common to both neurons (input
$S$). The binary stochastic signal $\eta(t)$ has value $A$ with
probability $q$ and $0$ otherwise, drawn independently for successive
time bins. Background activity is modeled by independent Gaussian
white noise in both scenarios. The input $G$ corresponds to the simplified
model presented in \cite[cf. Fig.4]{DeLaRocha07_802} that explains
the dependence of the correlation transmission of the firing rate.
In order to exclude this dependence, throughout \prettyref{fig:7}
we choose the parameters such that the mean activity of the neurons
remains unchanged. As shown in the marginal distribution if the input
current to a single neuron in \prettyref{fig:7}B, in the scenario
$S$ the binary process $\eta$ causes an additional peak with weight
$q$ centered around $A$. Equal activity in both scenarios requires
a constant probability mass above threshold $\theta$, which can be
achieved by appropriate choice of $\sigma_{S}<\sigma_{G}$. In scenario
$G$ the input correlation equals the fraction of shared input $\rhoin=c$,
as in \cite{DeLaRocha07_802}, whereas in scenario $S$ the input
correlation is $\rhoin=\frac{\Var[\eta]}{\Var[\eta]+\sigma_{S}^{2}}$,
where $\Var[\eta]=q(1-q)A^{2}$ is the variance of the binary input
signal $\eta(t)$. Comparing both scenarios, in \prettyref{fig:7}C-G
we choose $q$ such that the same input correlation is realized.

As for our spiking model, \prettyref{fig:7}C shows an increased correlation
transmission due to input synchrony. This observation can be intuitively
understood from the joint probability distribution of the inputs (\prettyref{fig:7}D-G).
Whenever any of the inputs exceeds the threshold ($I_{1,2}>\theta$)
the corresponding neuron becomes active, whenever both inputs exceed
threshold at the same time ($I_{1}>\theta\,\wedge\, I_{2}>\theta$),
both neurons are synchronously active. Therefore, $\langle f_{1}\rangle=\int_{\theta}^{\infty}dI_{1}\int_{-\infty}^{\infty}\: dI_{2}\: p(I_{1},I_{2})$,
the probability mass on the right side of $\theta$ for input $I_{1}$
(corresponding definition for $\langle f_{2}\rangle$), is a measure
for the activity of the neurons. Analogously, $\langle f_{1}f_{2}\rangle=\int_{\theta}^{\infty}\int_{\theta}^{\infty}\: dI_{1}dI_{2}\: p(I_{1},I_{2})$,
the probability mass in the upper right quadrant above both thresholds
is a measure for the output correlation between both neurons. Since
by our model definition the mean activity of both neurons is kept
constant, the masses $\langle f_{1}\rangle$ and $\langle f_{2}\rangle$
are equal in all four cases. However, the decisive difference between
scenarios with inputs $G$ and $S$ is the proportion of $\langle f_{1}f_{2}\rangle$
on the total mass above threshold $\langle f_{1}\rangle=\langle f_{2}\rangle$.
This proportion is increased by the common synchronous events modeled
as $\eta$, observable by comparing \prettyref{fig:7}D,E. The more
this proportion approaches $1$, $\langle f_{1}f_{2}\rangle\simeq\langle f_{1}\rangle=\langle f_{2}\rangle$,
the more the activity of both neurons is driven by $\eta$ (\prettyref{fig:7}F).
At the same time the contribution of the disjoint fluctuations on
the output activity is more and more suppressed. As the correlation
coefficient relates the common to the total fluctuations, the correlation
between the outputs can exceed the input correlation if the transmission
of the common input becomes more reliable than the transmission of
the disjoint input (cf. marker F in \prettyref{fig:7}C).

The situation illustrated in \prettyref{fig:7} is a caricature of
signal transmission by a pair of neurons of a cell assembly. The signal
of interest among the members of the assembly consists of synchronously
arriving synaptic events from peer neurons of the same assembly. In
our toy model such a volley is represented by an impulse of large
amplitude $A$. The remaining inputs are functionally considered as
noise and cause the dispersion of $I_{1}$ and $I_{2}$ observable
in \prettyref{fig:7}D-F. In the regime of sufficiently high synchrony
(corresponding to large $A$) in \prettyref{fig:7}F, the noise alone
rarely causes the neurons to be activated, it is suppressed in the
output signal due to the threshold. The synchrony coded signal, however,
reliably activates both neurons, moving $I_{1}$ and $I_{2}$ into
the upper right quadrant. Thus a synchronous volley is always mapped
to $1$ in the output, irrespective of the fluctuations caused by
the noise. In short, the non-linearity of neurons suppresses the noise
in the input while reliably detecting and transmitting the signal.
A similar effect of noise cancellation has recently been described
to prolong memory life-time in chain-like feed forward structures
\cite{Toyoizumi12_pre}.

\subsection{Limitations and Possible Extensions }

Several aspects of this study need to be taken into account when relating
the results to other studies and to biological systems. The multiple
interaction process as a model for correlated neural activity might
seem unrealistic at first sight. However, a similar correlation structure
can easily be obtained from the activity of a population of $N$ neurons.
Imagine each of the neurons to receive a set of uncorrelated afferents
causing a certain mean membrane potential $\mu$ and variance $\sigma^{2}$.
The entire population is then described by a membrane potential distribution
$p(V)$. In addition, each neuron receives a synaptic input of amplitude
$w$ that is common to all neurons. Whenever this input carries a
synaptic impulse, each of the $N$ neurons in the population has a
certain probability to emit a spike in direct response. The probability
equals the amount of density shifted across threshold by the common
synaptic event. Given the value $p(V_{\theta})$ and its slope \textrm{$\frac{\partial p}{\partial V}(V_{\theta})$}
of the membrane potential density at threshold \textrm{$V_{\theta}$,
the response probability is $P_{\mathrm{inst}}(w)=wp(V_{\theta})-\frac{1}{2}w^{2}\frac{\partial p}{\partial V}(V_{\theta})+O(w^{3})$
to second order in the synaptic weight $w$.} Employing the diffusion
approximation to the leaky integrate-and-fire neuron, the density
vanishes at threshold $p(V_{\theta})=0$ and the slope is given by
$\frac{\partial p}{\partial V}(V_{\theta})=-\frac{\nu_{\out}\taum}{2\sigma^{2}}$
\cite{Brunel00_183}. The response probability hence is $P_{\mathrm{inst}}(w)=\frac{\nu_{\out}\tau_{m}}{4\sigma^{2}}w^{2}$.
For typical values of $\nu_{\out}=20\Hz$, $\sigma=4\mV$, and $\taum=10\ms$
the estimate yields $w=5.7\mV$ to get the copy probability $p=0.1$
used in the current study. Such a synaptic amplitude is well in the
reported range for cortical connections \cite{DeWeese06_12206}. As
each of the neurons within the population responds independently,
the resulting distribution of the elicited response spikes is binomial,
as assumed by the MIP. Moreover, since our theory builds on top of
the moments of the complexity distribution it can be extended to other
processes introducing higher order spike correlations \cite{Stroeve01,Kuhn03_67}.

The boost of output correlation by synchronous synaptic impulses relies
on fast positive transients of the membrane potential and strong departures
from the stationary state: An incoming packet of synaptic impulses
brings the membrane potential over the threshold within short time.
Qualitatively, we therefore expect similar results for short, but
non-zero rise times of the synaptic currents. For long synaptic time
constants compared to the neuronal dynamics, however, the instantaneous
firing intensity follows the modulation of the synaptic current adiabatically
\cite{MorenoBote08,MorenoBote10_1528}. A similar increase of output
synchrony in this case can only be achieved if the static $f-I$ curve
of the neuron has a significant convex non-linearity.

The choice of the correlation measure is of importance when analyzing
spike correlations. It has been pointed out recently that the time
scale $\tau$ on which spike correlations are measured is among the
factors that can quantitatively change the results \cite{Cohen11_811}.
In particular, spike count correlations computed for time bins larger
than the intrinsic time scale of spike synchrony can be an ambiguous
estimate of input cross correlations \cite{Tchumatchenko10}. Considering
the exactly synchronous arrival of input events generated by the MIP,
we chose to measure count correlations on a small time scale $\tau=1\;\mathrm{ms}$
(except for the analysis regarding the limit of low input correlation
where we chose $\tau=100\;\mathrm{ms}$). The observation that the
difference between cross-correlation functions in absence and presence
of input synchrony is localized to their peak values and is otherwise
negligible reveals that this measure of zero-lag correlation fully
captures synchrony-induced correlations.

\subsection{Conclusion}

It had been proposed that the coordinated firing of cell assemblies
provides a means for the binding of coherent stimulus features \cite{Malsburg81,Bienenstock95,Singer95}.
Member neurons of such functional assemblies are interpreted to encode
the relevant information by synchronizing their spiking activity.
Under this assumption the spike synchrony produced by the assembly
can be considered as the signal and the remaining stochastic activity
as background noise. In order for a downstream neuron to reliably
convey and process the incoming signal received from the assembly,
it is essential to detect the synchronous input events carrying the
signal and to discern them from corrupting noise. Moreover, the processing
of such a synchrony-based code must occur independently of the firing
rate of the assembly members. We have shown that indeed the presence
of afferent spike synchrony leads to increased correlation susceptibility
compared to the transmission of shared input correlations. This finding,
that the correlation susceptibility is not a function of the firing
rate alone, demonstrates a limitation of the existing Gaussian white
noise theory that fails to explain the qualitatively different correlation
transmission due to synchrony \cite{DeLaRocha07_802}. We furthermore
have shown that under realistic conditions cortical neurons are able
to correlate their output stronger than the correlation they receive
in their input, yielding a correlation transmission coefficient $>1$.
This observation is also accompanied by non-changing firing rates.
This finding extends the prevailing view of neural correlations two-fold:
The correlation susceptibility exceeding unity invalidates the description
of correlation propagation as a 'transmission' per se. And moreover,
the dependence of the output correlation on the type of the input
correlation, not on its magnitude alone, demands an extended definition
of a correlation transmission coefficient. We have shown in a mechanistic
model how this coefficient exceeding unity results from the non-linearity
of cortical neurons enabling them to actively suppress the noise in
their input, thus sharpening the signal and improving the signal-to-noise
ratio. This observation is in agreement with the mechanism of synfire
activity where pulses of synchronized activity travel through feed
forward structures in a stable manner \cite{Diesmann99}. From our
findings we conclude that the boosting of correlation transmission
renders input synchrony highly effective compared to shared input
in causing closely time-locked output spikes in a task dependent and
time modulated manner, as observed in vivo \cite{Kilavik09_12653}.

\noindent 
\end{document}